\documentclass[sigconf]{acmart}
\usepackage{graphicx}
\usepackage{amsmath}
\usepackage{booktabs}
\usepackage{algorithm}
\usepackage{algorithmic}
\usepackage{amsfonts}
\usepackage{multirow}
\usepackage{makecell}
\usepackage{subfigure}
\usepackage{color}
\usepackage{bm}
\usepackage{epstopdf}
\usepackage{url}
\usepackage[cal=cm]{mathalfa}
\usepackage{balance}
\usepackage{threeparttable}
\usepackage{lipsum}
\usepackage{enumitem}
 \usepackage{pifont}
 \usepackage{tabularx}
 \usepackage{makecell}
 \usepackage{float}

\AtBeginDocument{%
  \providecommand\BibTeX{{%
    \normalfont B\kern-0.5em{\scshape i\kern-0.25em b}\kern-0.8em\TeX}}}

\renewcommand\footnotetextcopyrightpermission[1]{}

\begin{document}
\title{ChorusCVR: Chorus Supervision for Entire Space Post-Click Conversion Rate Modeling}

\author{Wei Cheng$^\S$}
\affiliation{
  \institution{Kuaishou Technology}
  \country{chengwei07@kuaishou.com}
 }
 
\author{Yucheng Lu$^\S$}
\affiliation{
  \institution{Kuaishou Technology}
  \country{luyucheng@kuaishou.com}
 }

\author{Boyang Xia$^\S$}
\affiliation{
  \institution{Kuaishou Technology}
  \country{xiaboyang@kuaishou.com}
 }
 
\author{Jiangxia Cao$^{\star}$}
\thanks{$^\S$Equal contributions to this work}
\thanks{$^\star$Corresponding author.}
\affiliation{
  \institution{Kuaishou Technology}
  \country{caojiangxia@kuaishou.com}
 }

 \author{Kuan Xu}
\affiliation{
  \institution{Kuaishou Technology}
  \country{xukuan@kuaishou.com}
 }

 \author{Mingxing Wen}
\affiliation{
  \institution{Kuaishou Technology}
  \country{wenmingxing@kuaishou.com}
 }

\author{Wei Jiang}
\affiliation{
  \institution{Kuaishou Technology}
  \country{jiangwei@kuaishou.com}
 }

 \author{Jiaming Zhang}
\affiliation{
  \institution{Kuaishou Technology}
  \country{zhangjiaming07@kuaishou.com}
 }
 
  \author{Zhaojie Liu}
\affiliation{
  \institution{Kuaishou Technology}
  \country{zhaotianxing@kuaishou.com}
 }

 \author{Liyin Hong}
\affiliation{
  \institution{Kuaishou Technology}
  \country{hongliyin@kuaishou.com}
 }

 \author{Kun Gai}
\affiliation{
  \institution{Unaffiliated}
  \country{gai.kun@qq.com}
 }

 \author{Guorui Zhou}
\affiliation{
  \institution{Kuaishou Technology}
  \country{zhouguorui@kuaishou.com}
 }

\renewcommand{\shorttitle}{ChorusNet}

\begin{abstract}
Post-click conversion rate (CVR) estimation is a vital task in many recommender systems of revenue businesses, \emph{e.g.,} e-commerce and advertising. In a perspective of sample, a typical CVR positive sample usually goes through a funnel of \textit{exposure}$\to$\textit{click}$\to$\textit{conversion}. For lack of post-event labels for un-clicked samples, CVR learning task commonly only utilizes clicked samples, rather than all exposed samples as for click-through rate (CTR) learning task. However, during online inference, CVR and CTR are estimated on the same assumed exposure space, which leads to a inconsistency of sample space between training and inference, \emph{i.e.,} sample selection bias (SSB). To alleviate SSB, previous wisdom proposes to design novel auxiliary tasks to enable the CVR learning on un-click training samples, such as CTCVR and counterfactual CVR, \emph{etc}. Although alleviating SSB to some extent, none of them pay attention to the discrimination between ambiguous negative samples (un-clicked) and factual negative samples (clicked but un-converted) during modelling, which makes CVR model lacks robustness. To full this gap, we propose a novel \textbf{ChorusCVR} model to realize debiased CVR learning in entire-space. We propose a Negative sample Discrimination Module (NDM), which aims to provide robust soft labels with the ability to discriminate factual negative samples (clicked but un-converted) from ambiguous negative samples (un-clicked). Moreover, we propose a  Soft Alignment Module (SAM) to supervise CVR learning with several alignment objectives using generated soft labels. 
Extensive offline experiments and online A/B testing at Kuaishou’s e-commerce live service validates the efficacy of our ChorusCVR. 
\end{abstract}



\maketitle

\section{Introduction}

\begin{figure}[t]
  \centering
  \includegraphics[width=7cm]{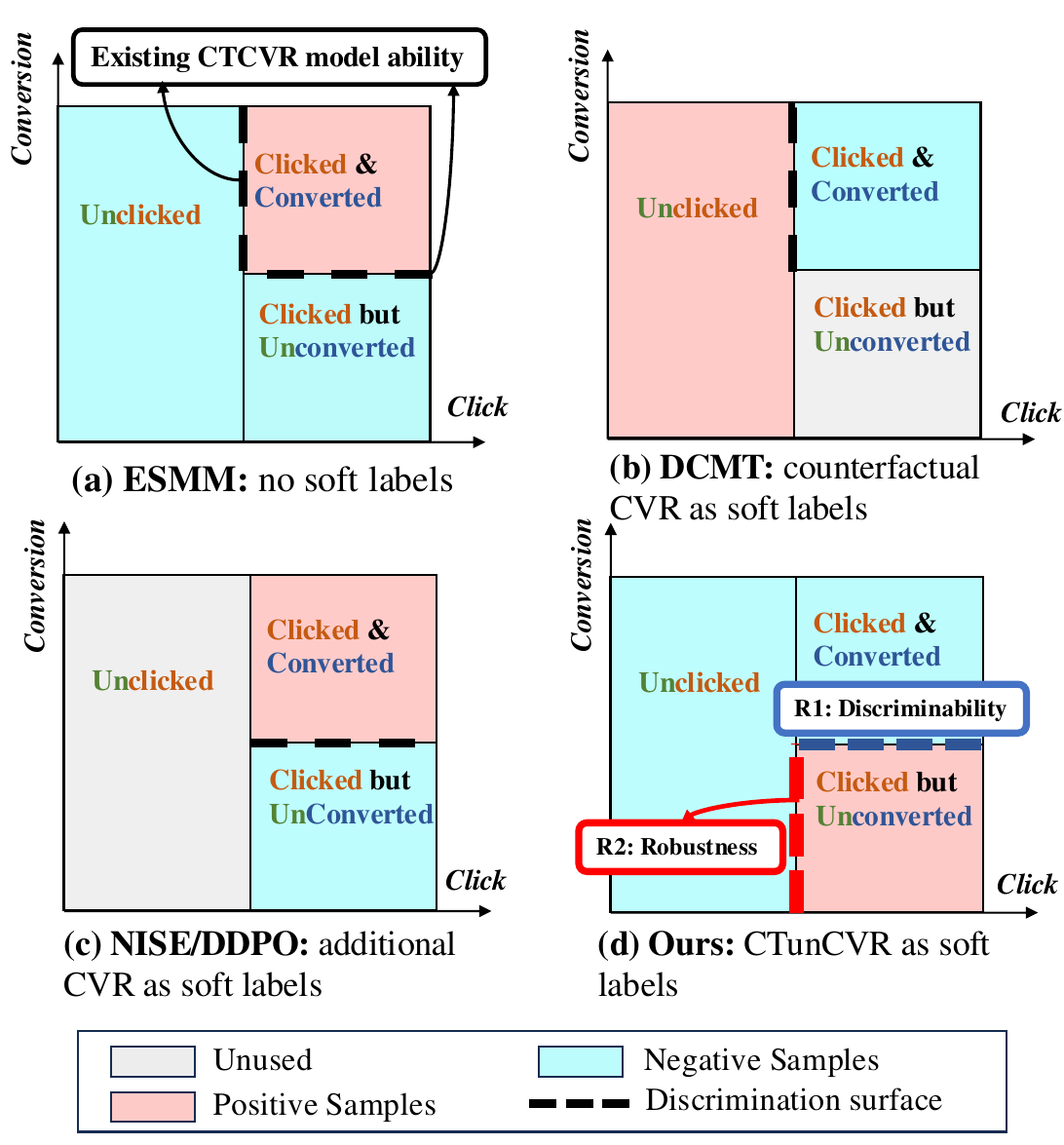}
  \vspace{-1em}
  \caption{A conceptual comparison between our proposed ChorusCVR and existing CVR models on the perspective of the discrimination spaces of soft labels.}
  \label{intro}
  \vspace{-2em}
\end{figure}
Recommender systems are crafted to provide users with personalized content (videos, products and ads, \emph{etc.,}) that match their preferences \cite{youtubenet,sim,mmoe,din}. Generally, industrial RecSys typically divided into two major stages. 1) Retrieval stage, which aims to search thousands of related candidates from massive item pool. 2) Ranking stage, which aims to estimate interaction probability, \emph{e.g.,} click-through rate (CTR) and post-click conversion rate (CVR), for each user-item pair for retrieved candidates, and select a set of best items for users. In this paper, we focus on the post-click conversion rate (CVR) estimation task during ranking stage.

\textit{Problem statement.} Typically, a positive CVR sample follows the following data funnel: 
\textit{exposure} $\mathcal{D}$$\to$\textit{click} $\mathcal{O}$$\to$\textit{conversion} $\mathcal{R}$, where the \textit{click} space $\mathcal{O}$ is around about $4\sim6\%$ of \textit{exposure} space $\mathcal{D}$ and the \textit{conversion} space $\mathcal{R}$ takes up $2\sim4\%$ of \textit{click} space $\mathcal{O}$.
Different with CTR which is learned using exposure space samples, CVR is typically learned using only click space samples because we are unaware the un-clicked samples would be converted or not. 
However, during online inference, the CTR and CVR scores are utilized in the same assumed exposure space, which leads to a well-known mismatch sample selected bias (SSB) issue \cite{ssb1,ssb2,ssb3,dcmt}, that CVR learning module is trained in \textit{click} space $\mathcal{O}$ but is used for inference at  \textit{exposure} space $\mathcal{D}$.

\textit{Motivation.} To alleviate the SSB problem, previous wisdom introduce several techniques to extend CVR task to \textit{exposure} space.
Specifically, ESMM \cite{essm} propose a click-through \& conversion rate (CTCVR) task to merge two CVR and CTR scores as one score to supervised it in \textit{exposure} space, which successfully extend the CVR to entire space to solve space inconsistency between training and inference.
Unfortunately, the CTCVR loss made a strong assumption that \textbf{\textbf{un-clicked} training samples are hard negative samples in CTCVR training}. This assumption overlooks some ambiguous negative samples which may be easy for users to buy after he/she clicked, but without chance to be clicked yet \cite{multiipw,dcmt}.
To alleviate this false negative sample issue, the recent works are dedicated to find reasonable pseudo soft labels to for \textbf{un-clicked} sample learning.
Specifically, the DCMT \cite{dcmt} propose to regularize the CVR objectives by a complementary constraint with a novel counterfactual CVR objective. 
For counterfactual CVR learning, DCMT first assumes all un-clicked items as positive samples while all converted items are negative samples, and then apply a $CVR = 1 - counterfactualCVR$ constraints for CVR learning module, as shown in Figure~\ref{intro}(b). 
Besides, the NISE \cite{nise} and DDPO \cite{ddpo} first utilize the outputs of an additional CVR tower learned in click space to act as pseudo soft label, and then employ a cross-entropy constraints $CVR\thickapprox extraCVR$ in un-click space, as shown in Figure~\ref{intro}(c).

It has come to our attention that the quality of soft CVR labels is the key to mitigating SSB issues.  So we ask, \textit{what requirements should an ideal soft CVR label satisfies}? Our key insight is, an ideal soft label should at least satisfy two requirements: \textbf{R1. Discriminability}: for a clicked sample, the label can discriminate it would be converted or not; \textbf{R2. Robustness}: for un-converted sample, the label can separate the factually un-converted sample in click space from those ambiguous un-converted sample in un-click space. With this in mind, we  present the discrimination surface of soft labels in existing methods (DCMT, NISE and DDPO) in Figure~\ref{intro}. We find their discrimination surface are fully overlapped with certain part of the surfaces of  ESMM (CTCVR task w/o soft labels), which miss either \textbf{R1} or \textbf{R2}. Thus, few of existing methods can meet all these requirements.

To fill this gap, we present a novel entire-space dual multi-task learning model, namly \textbf{ChorusCVR}, to realize effective CVR learning in un-click space that  fulfills both \textbf{R1} and \textbf{R2} (see Figure~\ref{intro}(d)).  The ChorusCVR consists of two modules, \emph{i.e.,} \textbf{N}egative sample \textbf{D}iscrimination \textbf{M}odule (NDM) and \textbf{S}oft \textbf{A}lignment \textbf{M}odule (SAM). In NDM, we introduce a novel CTunCVR auxiliary task, to provide robust soft CVR labels with the ability to discriminate factual CVR negative samples (clicked but un-converted) and ambiguous CVR negative samples (un-clicked).  In SAM, we utilize generated CTunCVR soft outputs to supervise  CVR learning with several alignment objectives, to realize debiased CVR learning in entire-space. Our contributions can be summarized as follows:
\begin{itemize}[leftmargin=*,align=left]
    \item We introduce a novel CTunCVR auxiliary task to provide soft CVR labels with both discriminability and robustness in entire space.
    \item We propose a novel ChorusCVR model with effective alignment objectives for debiased CVR modelling in entire space.
    \item We conduct extensive experiments on both public and production environment datasets and online A/B testing to verify the efficacy of our method, which show that our ChorusCVR achieves superior performance over all existing state-of-the-art methods.  
\end{itemize}
\section{Methodology}

\begin{figure*}[t]
  \centering
\includegraphics[width=0.70\textwidth]{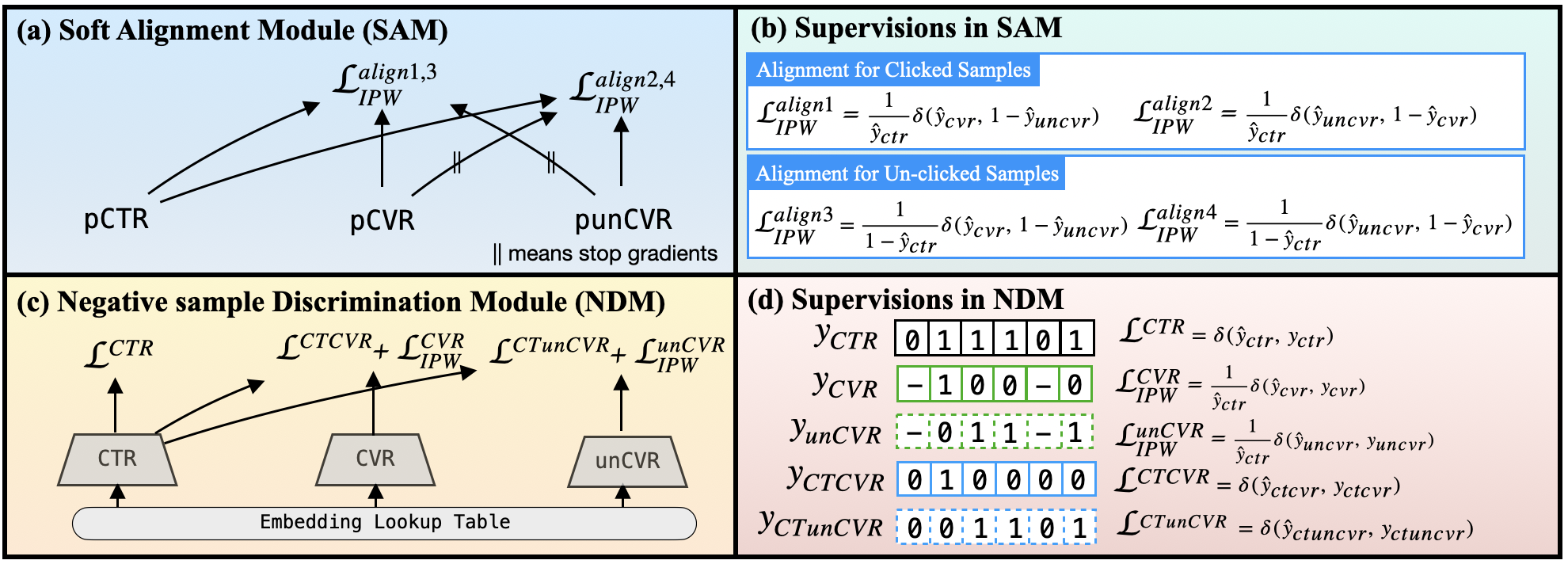}
  \vspace{-1em}
  \caption{Systematic overview of our Chorus CVR model.}
  \label{choruscvr}
  \vspace{-1em}
\end{figure*}

\subsection{Preliminary}
In the ranking stage of industrial recommendation system, all \textit{exposure} user-item pairs will be collected and formed as a data-streaming for model training, i.e., $\mathcal{D}$.
Specifically, each user-item sample in $\mathcal{D}$ could represent as $(u, i, \{\mathbf{x}_u, \mathbf{x}_i,$ $\mathbf{x}_{ui}\}, o_{ui}, r_{ui}) \in \mathcal{D}$, where $u$/$i$ denotes the user-item pair, $\mathbf{x}_u\in\mathbb{R}^{d_u}, \mathbf{x}_i\in\mathbb{R}^{d_i}, \mathbf{x}_{ui}\in\mathbb{R}^{d_{ui}}$ are the user-side features (e.g., user ID), item-side features (e.g., item ID), and item-aware cross features (e.g., SIM \cite{sim}).
The $o_{ui}\in\{0,1\}$ and $r_{ui}\in\{0,1\}$ are user-item ground-truth interacted labels, where $o_{ui}$ denotes whether user $u$ clicked item $i$ and $r_{ui}$ denotes whether user $u$ converted item $i$. 
According to the entire \textbf{\textit{exposure} space} $\mathcal{D}$, we could further obtain several subset spaces:
\begin{itemize}[leftmargin=*,align=left]
\item \textbf{\textit{Click} space} $\mathcal{O}\in\mathcal{D}$, if click label $o_{ui} = 1$.
\item \textbf{\textit{un-Click} space} $\mathcal{N}=\mathcal{D} - \mathcal{O}$, if click label $o_{ui} = 0$.
\item \textbf{\textit{Conversion} space} $\mathcal{R}\in\mathcal{O}$: if label $o_{ui}=1$ and $r_{ui}=1$.
\item \textbf{\textit{un-Conversion} space} $\mathcal{M}=\mathcal{O}-\mathcal{R}$: if label $o_{ui}=1$ and $r_{ui}=0$.
\end{itemize}
Based on them, a simple ranking model can be formed as:
\begin{equation}
\begin{split}
&\hat{y}^{ctr}_{ui} = \texttt{MLP}^{ctr}(\mathbf{x}_{ui}),\quad \hat{y}^{cvr}_{ui} = \texttt{MLP}^{cvr}(\mathbf{x}_{ui}),\\
&\mathbf{x}_{ui} = \texttt{Multi-Task-Encoder}(\mathbf{x}_u\oplus \mathbf{x}_i\oplus \mathbf{x}_{ui}),\\
\end{split}
\label{base}
\end{equation}
where the $\oplus$ denotes the concatenate operator, $\mathbf{x}\in\mathbb{R}^d$ is the encoded hidden states, and $\texttt{MLP}(\cdot)$ denotes a stacked neural-network. We use a share-bottom based multi-task paradigm to predict CTR and CVR scores, $\hat{y}^{ctr},\hat{y}^{cvr}$.
Next, we directly minimize the cross-entropy binary classification loss to train CTR tower and CVR towers with corresponding space samples:
\begin{equation}
\begin{split}
&\mathcal{L}^{ctr} = - \frac{1}{|\mathcal{D}|}\big(\sum_{(u,i)\in\mathcal{D}}\delta(\hat{y}^{ctr}_{ui}, o_{ui})\big),\\
&\mathcal{L}^{cvr} = - \frac{1}{|\mathcal{O}|}\big(\sum_{(u,i)\in\mathcal{O}}\delta(\hat{y}^{cvr}_{ui}, r_{ui})\big).
\end{split}
\label{crossentropy}
\end{equation}
In inference, given the hundreds item candidates in a certain user request, we could obtain predicted CTCVR by  $\hat{y}^{ctcvr}_{ui} =\hat{y}^{ctr}_{ui} \cdot \hat{y}^{cvr}_{ui}$ for each item, which is used for final ranking. Then top K highest items will be returned and shown to user. The CVR are learned in click space during training but be predicted in an assumed explore space during inference, which brings up the question of sample selection bias problem.

To alleviate sample selection bias,  ESMM \cite{essm} expand the click-space CVR learning task to exposure-space CTCVR learning task, to directly solve the inconsistency between training and inference:
\begin{equation}
\begin{split}
\mathcal{L}^{ctcvr} = &- \frac{1}{|\mathcal{D}|}\Big(\sum_{(u,i)\in\mathcal{D}}\delta(\hat{y}^{ctr}_{ui}\cdot\hat{y}^{cvr}_{ui}, o_{ui}\cdot r_{ui})\Big)
\end{split}
\label{ctxcvr}
\end{equation}
which treats all un-clicked samples as negative samples of CTCVR task. However those un-clicked samples that would be converted if clicked, which are falsely negative samples,  still leads to missing not at random (MNAR) problem  \cite{multiipw}. To mitigate this problem, inverse propensity weighting (IPW) \cite{multiipw,escm2} based method inversely weight the CVR loss in click space by propensity score of observing  $(u,i)$ in click space $\mathcal{O}$, to eliminate the influence of click event to CVR estimation in entire space $D$
\begin{equation}
\small
\begin{split}
\mathcal{L}^{cvr}_{IPW} =& - \frac{1}{|\mathcal{O}|}\Big(\sum_{(u,i)\in\mathcal{O}}\frac{\delta(\hat{y}^{cvr}_{ui}, r_{ui})}{\hat{y}^{ctr}_{ui}}\Big),
\end{split}
\label{cvripw}
\end{equation}
Our method is based on above ESMM with IPW  framework. Although alleviating SSB and MNAR problem, IPW-based methods still lack reasonable labels for \textit{un-clicked} samples, which we solve by generating discriminative and robust soft labels.

\subsection{ChorusCVR}
In this section, we dive into ChorusCVR and explain how we realize entire-space debiased CVR learning by generating discriminative and robust soft CVR labels (as shown in Figure~\ref{choruscvr}).

\subsubsection{Negative sample Discrimination Module (NDM)}

As mentioned before, the soft labels introduced by previous works are suboptimal for lack either discriminability or robustness. As shown in Figure~\ref{intro} (d), an ideal discrimination surface should separate the factual negative samples (clicked but un-converted) from positive samples (clicked \& converted), and factual negative samples from ambiguous negative samples (un-clicked). With this in mind, we find the ideal discrimination surface implies a new task, CTunCVR prediction. We formulate CTunCVR labels as:
\begin{equation}
y^{ctuncvr} = o_{ui} * (1-r_{ui}) = 
\begin{cases} 
1 & o_{ui}=1~\&~r_{ui} = 0, \\
0 &  o_{ui}=0, \\
0 & o_{ui}=1~\&~r_{ui} = 1,
\end{cases}
\end{equation}
where only \textit{clicked but un-converted} samples are positive samples, both \textit{clicked \& converted} and \textit{un-clicked} samples are negative samples. Instead of directly predicting CTunCVR score in exposure space, we follow a typical two-stage prediction paradigm to obtain CTunCVR to 
reduce cumulative error. We firstly introduce an additional unCVR tower to predict unCVR score $\hat{y}^{uncvr}$, then combine it with $\hat{y}^{ctr}$ to form CTunCVR score:
\begin{equation}
\begin{split}
\hat{y}^{uncvr}_{ui} = \hat{y}^{ctr}_{ui}\cdot\hat{y}^{cvr}_{ui}\quad \quad
\hat{y}^{ctuncvr}_{ui} =\hat{y}^{ctr}_{ui}\cdot\hat{y}^{uncvr}_{ui}
\end{split}
\label{uncvr}
\end{equation}
Then we can naturally optimize CTunCVR objective in exposure space by cross entropy loss: 
\begin{equation}
\begin{split}
\mathcal{L}^{ctuncvr} = - \frac{1}{|\mathcal{D}|}\Big(\sum_{(u,i)\in\mathcal{D}}\delta(\hat{y}^{ctuncvr}_{ui}, o_{ui} * (1-r_{ui})\Big).
\end{split}
\label{uncvr}
\end{equation}
With the help of $\mathcal{L}^{ctuncvr}$ and an extra \textbf{unCVR prediction result} $\hat{y}^{uncvr}_{ui}$, we can narrow down the aforementioned problem to consider \textbf{R1. Discriminability} and \textbf{R2. Robustness} problem at same time.
For the $\hat{y}^{uncvr}_{ui}$ generation, we add an mirror unCVR tower which similar with the Eq.(\ref{base}) and (\ref{crossentropy}):
%
%
%
%
%
%
%
\begin{equation}
\begin{split}
\hat{y}^{uncvr}_{ui} &= \texttt{MLP}^{uncvr}(\mathbf{x}_{ui}), \\
\mathcal{L}^{uncvr} = - \frac{1}{|\mathcal{O}|}\Big(&\sum_{(u,i)\in\mathcal{O}}\delta\big(\hat{y}^{uncvr}_{ui}, 1-r_{ui})\big)\Big)
\end{split}
\label{uncvr}
\end{equation}
Next, analogously with the Eq.(\ref{cvripw}), we then adopt the predicted click $\hat{y}^{ctr}_{ui}$ to inversely weight the unCVR error, to $\mathcal{L}^{uncvr}$ as:
\begin{equation}
\begin{split}
\mathcal{L}^{uncvr}_{IPW} = - \frac{1}
{|\mathcal{O}|}\Big(\sum_{(u,i)\in\mathcal{O}}\frac{\delta(\hat{y}^{uncvr}_{ui}, 1-r_{ui})}{\hat{y}^{ctr}_{ui}}\Big)
\end{split}
\label{uncvr}
\end{equation}
In this way, the \textit{click} space tendency can be alleviated that higher/lower $pCTR$ sample will declined/enhanced for a fair training. So far we obtain debiased unCVR soft labels, which we will utilize to help the CTunCVR training and CVR component supervision.

\subsubsection{Soft Alignment Module (SAM)}
Up to now, we fulfill the initial goal of obtain high-quality soft labels in un-clicked space. In this section we present the solution to utilize the unCVR score as soft labels to supervise CVR learning, which we call \emph{soft alignment mechanism}. We first use $1-unCVR$ manner as soft labels for entropy-based CVR learning. In the same time, we also use $1-CVR$ manner to generate soft labels for unCVR learning, in a mutual supervision fashion to align unCVR predictions to CVR. All these objectives are inversely weighted by predicted CTR in a IPW paradigm (see $\mathcal{L}^{align1}_{IPW}$ and $\mathcal{L}^{align2}_{IPW}$ in Figure.~\ref{choruscvr}). To further alleviate SSB for un-click space, we also propose a \textit{un-click space IPW} approach, to inversely weight the un-click samples with $1-pCTR$ for CTR and unCVR alignment objectives (see $\mathcal{L}^{align3}_{IPW}$ and $\mathcal{L}^{align4}_{IPW}$ in Figure.~\ref{choruscvr}). Overall, all alignment objectives are as follows:
\begin{equation}
\begin{split}
\mathcal{L}^{align}_{IPW} = &- \frac{1}{|\mathcal{O}|}\big(\frac{\delta(\hat{y}^{cvr}_{ui}, 1-\texttt{sg}(\hat{y}^{uncvr}_{ui}))}{\hat{y}^{ctr}_{ui}}\big)
- \frac{1}{|\mathcal{N}|}\big(\frac{\delta(\hat{y}^{cvr}_{ui}, 1-\texttt{sg}(\hat{y}^{uncvr}_{ui}))}{1-\hat{y}^{ctr}_{ui}}\big)\\
&- \frac{1}{|\mathcal{O}|}\big(\frac{\delta(\hat{y}^{uncvr}_{ui}, 1-\texttt{sg}(\hat{y}^{cvr}_{ui}))}{\hat{y}^{ctr}_{ui}}\big)
- \frac{1}{|\mathcal{N}|}\big(\frac{\delta(\hat{y}^{uncvr}_{ui}, 1-\texttt{sg}(\hat{y}^{cvr}_{ui}))}{1-\hat{y}^{ctr}_{ui}}\big)
\end{split}
\label{soft}
\end{equation}

where the $\texttt{sg}(\cdot)$ means the stop gradient function, the $\hat{y}^{ctr}_{ui}, (1 - \hat{y}^{ctr}_{ui})$ denote the click propensity in the \textit{click} and \textit{un-click} space, respectively.
All losses of our ChorusCVR are as follows:
\begin{equation}
\begin{split}
\mathcal{L} = \mathcal{L}^{ctcvr} + \mathcal{L}^{cvr}_{IPW} + \mathcal{L}^{ctuncvr} + \mathcal{L}^{uncvr}_{IPW} + \mathcal{L}^{align}_{IPW}
\end{split}
\label{soft}
\end{equation}
In this way, our ChorusCVR  make CVR and unCVR supervise each other during training, which results in an equilibrium.

\begin{table}[t!]
\centering
\caption{Offline results(\%) in terms of CTR-AUC, CTCVR-AUC and logloss at Ali-CCP and Kuaishou.}
\vspace{-1em}
\setlength{\tabcolsep}{3pt}{
\begin{tabular}{l|cc|cc}
\toprule
\multirow{4}{*}{\makecell{Models}} 
& \multicolumn{2}{c|}{Ali-CCP} & \multicolumn{2}{c}{Kuaishou}   \\ 
\cmidrule(r){2-5} & \multicolumn{2}{c|}{AUC}  & \multicolumn{2}{c}{AUC}   \\ 
\cmidrule(r){2-3} \cmidrule(r){4-5} & CVR  & CTCVR &  CVR & CTCVR  \\
\hline
ESMM \cite{essm} & 0.5963 & 0.5802 & 0.8609 & 0.9276\\ 
ESCM$^2$-DR \cite{escm2} & 0.6354 & 0.6203& 0.8617 & 0.9280\\
ESCM$^2$-IPW \cite{escm2} & 0.6385 & 0.6126& 0.8619 & 0.9283\\
UKD \cite{ukd} & 0.6451 & 0.6282& 0.8615 & 0.9279\\
DCMT \cite{dcmt} & 0.6447 & \underline{0.6375} & \underline{0.8628} & 0.9281\\
DDPO \cite{ddpo} & \underline{0.6496} & 0.6326& 0.8623 & \underline{0.9289}\\
NISE \cite{nise} & 0.6418 & 0.6291 & 0.8614 & 0.9274\\
ChorusCVR & \textbf{0.6589} & \textbf{0.6401}& \textbf{0.8639} & \textbf{0.9304}\\
\hline
Rela.Impr. & +1.43\% & +0.40\% & +0.13\% & +0.16\% \\
\hline
ChorusCVR w/o NDM & 0.6498 & 0.6347& 0.8625 & 0.9284\\
ChorusCVR w/o SAM & 0.6407 & 0.6139 & 0.8622 & 0.9281\\
\bottomrule
\end{tabular}
}
\vspace{-2em}
\label{mainoffline}
\end{table}

\section{Experiments}
In this section, we conduct extensive online and offline experiments to verify the efficacy of our model following 3 research questions: 
\begin{itemize}[leftmargin=*,align=left]
\item \textbf{RQ1:} How does our model perform on industrial recommendation datasets?
\item \textbf{RQ2:} Can our model bring improvements of e-commerce A/B test metrics on online product environment?
\item \textbf{RQ3:} Can our model alleviate the bias on un-clicked samples?
\end{itemize}
\textbf{Datasets.} To evaluate the performance of our method and comparison baselines, we conduct comprehensive experiments on both public and industrial datasets.

• Public dataset: The Ali-CCP (Alibaba Click and Conversion Prediction) dataset \cite{essm} is a benchmark dataset for CVR and CTR prediction, which collected from traffic logs in Taobao e-commerce platform. Ali-CCP contains 33 features and 84M samples. The training set contains 42M exposure samples, 1.6M click samples and 9k conversion samples and the test set contains 42M exposure samples, 1.7M click samples and 9.4k conversion samples.

• Industrial dataset: The industrial dataset is colloected from Kuaishou e-commerce live-streaming platform. Kuaishou e-commerce live-streaming is a popular content interest e-commerce platform with \textbf{tens of millions} daily active users. The dataset contains more than 1000 features including user profiles (gender, age, etc), user actions (click, buy, etc), seller profiles (industry, sales, etc) and goods information. 

\textbf{Compared Methods.} ESMM \cite{essm} learns CVR task through a CTR task and a CTCVR task to alleviate ssb and data sparsity issues. ESCM$^2$-IPW \cite{escm2} incorporates the inverse propensity weighting (IPW) \cite{ipw} method to regularize ESMM’s CVR estimation. ESCM$^2$-DR \cite{escm2} augments ESCM$^2$-IPW with an auxiliary imputation loss to models the CVR with the Doubly Robust(DR) method. UKD \cite{ukd} introduces a transfer adversarial learning approach to generate soft conversion pseudo-labels in the unclicked space. DCMT \cite{dcmt} proposed a counterfactual mechanism to directly debias CVR in the entire space. DDPO \cite{ddpo} employs a conversion propensity prediction network to generate soft conversion pseudo-labels in the un-clicked space. NISE \cite{nise} follows a semi-supervised learning paradigm and use predicted CVR as CVR pseudo labels for un-clicked samples.

\textbf{Offline Results (RQ1):} 
We evaluate the efficacy of our model by the Area Under ROC (AUC) of CVR and CTCVR prediction tasks. The experimental results on two datasets are shown in Table 1. On Ali-CCP dataset, our model outperforms DDPO by 1.4\% (\textbf{0.6589} v.s. 0.6496) for CVR-AUC and DCMT by 0.4\% (\textbf{0.6401} v.s. 0.6375) for CTCVR-AUC, respectively. On Kuaishou dataset, our model outperforms DCMT by 0.1\% (\textbf{0.8639} v.s. 0.8628) for CVR-AUC and DDPO by 0.16\% (\textbf{0.9304} v.s. 0.9289) for CTCVR-AUC, respectively. In a nutshell, our model consistently outperforms the best-performing baselines on both two datasets in a large margin in terms of two tasks. 
\begin{figure}[t]
  \centering
  \includegraphics[width=5cm]{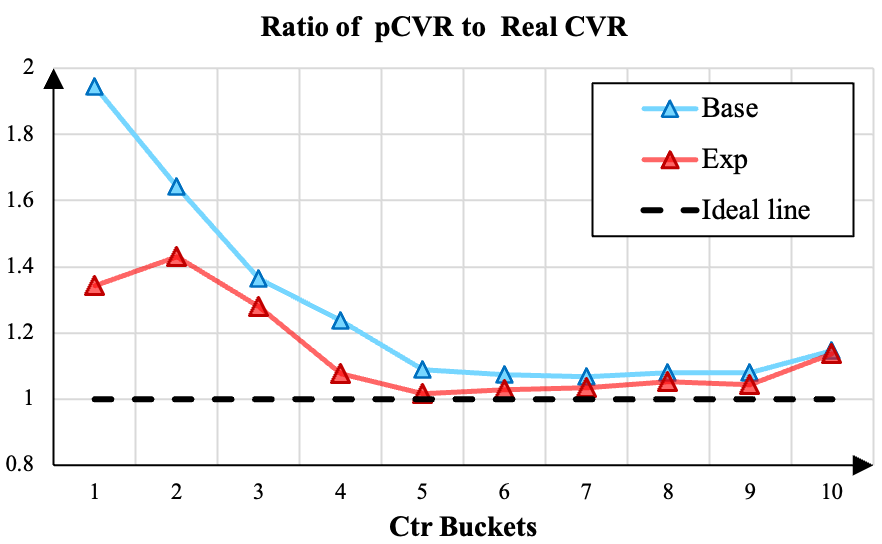}
  \vspace{-1em}
  \caption{The PCOC analysis.}
  \label{pcoc}
  \vspace{-2em}
\end{figure}
\textit{Ablation Study.} To verify the efficacy of each parts of ChorusCVR, we evaluate the performance of variants without NDM and SAM. It is noteworthy that the unCVR objective is optimized by 1 - $y^{CVR}$ label in click space in `ChorusCVR w/o NDM'. We find that without NDM, our model degrades 1.3\% (\textbf{0.6589} v.s. 0.6498) on CVR AUC on Ali-CCP dataset, for lack of discrimination between negative samples of different levels. Meanwhile, we can observe that without SAM, the `ChorusCVR w/o SAM' performs even poorer than `ChorusCVR w/o NDM', presents 4\% lower CTCVR AUC on Ali-CCP dataset, for lack of reasonable pesudo supervisions on un-click samples. 

\textbf{Online Results (RQ2):}
We deploy our method in production environment of Kuaishou e-commerce live-streaming platform to conduct online A/B testing for 8 days. 
Compared with the base model (DCMT), our model presents large improvements on CVR(+0.12\%), orders (+0.851\%) and DAC (+0.705\%) with 95\% confidence intervals.

\textbf{Analysis (RQ3):}
To investigate whether our method can address the bias in un-clicked samples, we try to compare the  accuracy of cvr predicitons between our method and baselines on un-clicked samples. However, the conversion labels of un-clicked samples are inaccessible in the real world. Inspired by the inverse propensity weighting based methods, we propose a compromised comparison approach, to use samples with low pCTR as a substitute for un-clicked samples and observe the model's pCVR on these low CTR samples. We show the estimated pCVR and actual cvr for samples with different pCVR the deviations between two curves signify the prediction bias of the model.

As shown in Fig. \ref{pcoc}, we can observe that DCMT represents obvious bias on low pCTR samples, which severely overestimates the CVR on low pCtr samples. However, our model accurately predicts the CVR on those low pCtr samples for reasonable CVR supervisions on un-clicked samples.

\section{Conclusion}
In this paper, to alleviate sample selection bias in CVR prediction task, we propose an effective method ChorusCVR. To generate discriminative and robust soft labels, we propose Negative sample Discrimination Module to obtain soft CTunCVR labels which can separate negative samples of different levels. Then we design a 
Soft Alignment Module for debiased CVR learning in un-click space with soft labels. We demonstrated the superior performance of the proposed
ChorusCVR in offline experiments. In addition, we conduct online A/B testing, obtaining +0.851\% improvements on orders of industrial e-commerce living stream, which demonstrates the effectiveness and universality of ChorusCVR in online
systems. 

\balance
\bibliographystyle{ACM-Reference-Format}
\bibliography{sample-base-extend.bib}


\begin{thebibliography}{15}


\ifx \showCODEN    \undefined \def \showCODEN     #1{\unskip}     \fi
\ifx \showDOI      \undefined \def \showDOI       #1{#1}\fi
\ifx \showISBNx    \undefined \def \showISBNx     #1{\unskip}     \fi
\ifx \showISBNxiii \undefined \def \showISBNxiii  #1{\unskip}     \fi
\ifx \showISSN     \undefined \def \showISSN      #1{\unskip}     \fi
\ifx \showLCCN     \undefined \def \showLCCN      #1{\unskip}     \fi
\ifx \shownote     \undefined \def \shownote      #1{#1}          \fi
\ifx \showarticletitle \undefined \def \showarticletitle #1{#1}   \fi
\ifx \showURL      \undefined \def \showURL       {\relax}        \fi
\providecommand\bibfield[2]{#2}
\providecommand\bibinfo[2]{#2}
\providecommand\natexlab[1]{#1}
\providecommand\showeprint[2][]{arXiv:#2}

\bibitem[Bareinboim et~al\mbox{.}(2022)]%
        {ssb3}
\bibfield{author}{\bibinfo{person}{Elias Bareinboim}, \bibinfo{person}{Jin Tian}, {and} \bibinfo{person}{Judea Pearl}.} \bibinfo{year}{2022}\natexlab{}.
\newblock \showarticletitle{Recovering from selection bias in causal and statistical inference}.
\newblock In \bibinfo{booktitle}{\emph{Probabilistic and causal inference: The works of Judea Pearl}}. \bibinfo{pages}{433--450}.
\newblock


\bibitem[Covington et~al\mbox{.}(2016)]%
        {youtubenet}
\bibfield{author}{\bibinfo{person}{Paul Covington}, \bibinfo{person}{Jay Adams}, {and} \bibinfo{person}{Emre Sargin}.} \bibinfo{year}{2016}\natexlab{}.
\newblock \showarticletitle{Deep neural networks for youtube recommendations}. In \bibinfo{booktitle}{\emph{Proceedings of the 10th ACM conference on recommender systems}}. \bibinfo{pages}{191--198}.
\newblock


\bibitem[De~Myttenaere et~al\mbox{.}(2014)]%
        {ssb1}
\bibfield{author}{\bibinfo{person}{Arnaud De~Myttenaere}, \bibinfo{person}{B{\'e}n{\'e}dicte~Le Grand}, \bibinfo{person}{Boris Golden}, {and} \bibinfo{person}{Fabrice Rossi}.} \bibinfo{year}{2014}\natexlab{}.
\newblock \showarticletitle{Reducing offline evaluation bias in recommendation systems}.
\newblock \bibinfo{journal}{\emph{arXiv preprint arXiv:1407.0822}} (\bibinfo{year}{2014}).
\newblock


\bibitem[Farajtabar et~al\mbox{.}(2018)]%
        {ssb2}
\bibfield{author}{\bibinfo{person}{Mehrdad Farajtabar}, \bibinfo{person}{Yinlam Chow}, {and} \bibinfo{person}{Mohammad Ghavamzadeh}.} \bibinfo{year}{2018}\natexlab{}.
\newblock \showarticletitle{More robust doubly robust off-policy evaluation}. In \bibinfo{booktitle}{\emph{International Conference on Machine Learning}}. PMLR, \bibinfo{pages}{1447--1456}.
\newblock


\bibitem[Huang et~al\mbox{.}(2024)]%
        {nise}
\bibfield{author}{\bibinfo{person}{Jiahui Huang}, \bibinfo{person}{Lan Zhang}, \bibinfo{person}{Junhao Wang}, \bibinfo{person}{Shanyang Jiang}, \bibinfo{person}{Dongbo Huang}, \bibinfo{person}{Cheng Ding}, {and} \bibinfo{person}{Lan Xu}.} \bibinfo{year}{2024}\natexlab{}.
\newblock \showarticletitle{Utilizing Non-click Samples via Semi-supervised Learning for Conversion Rate Prediction}. In \bibinfo{booktitle}{\emph{Proceedings of the 18th ACM Conference on Recommender Systems}}. \bibinfo{pages}{350--359}.
\newblock


\bibitem[Ma et~al\mbox{.}(2018b)]%
        {mmoe}
\bibfield{author}{\bibinfo{person}{Jiaqi Ma}, \bibinfo{person}{Zhe Zhao}, \bibinfo{person}{Xinyang Yi}, \bibinfo{person}{Jilin Chen}, \bibinfo{person}{Lichan Hong}, {and} \bibinfo{person}{Ed~H Chi}.} \bibinfo{year}{2018}\natexlab{b}.
\newblock \showarticletitle{Modeling Task Relationships in Multi-task Learning with Multi-gate Mixture-of-Experts}. In \bibinfo{booktitle}{\emph{ACM SIGKDD Conference on Knowledge Discovery and Data Mining (KDD)}}.
\newblock


\bibitem[Ma et~al\mbox{.}(2018a)]%
        {essm}
\bibfield{author}{\bibinfo{person}{Xiao Ma}, \bibinfo{person}{Liqin Zhao}, \bibinfo{person}{Guan Huang}, \bibinfo{person}{Zhi Wang}, \bibinfo{person}{Zelin Hu}, \bibinfo{person}{Xiaoqiang Zhu}, {and} \bibinfo{person}{Kun Gai}.} \bibinfo{year}{2018}\natexlab{a}.
\newblock \showarticletitle{Entire Space Multi-Task Model: An Effective Approach for Estimating Post-Click Conversion Rate}. In \bibinfo{booktitle}{\emph{International ACM SIGIR Conference on Research and Development in Information Retrieval (SIGIR)}}.
\newblock


\bibitem[Pi et~al\mbox{.}(2020)]%
        {sim}
\bibfield{author}{\bibinfo{person}{Qi Pi}, \bibinfo{person}{Guorui Zhou}, \bibinfo{person}{Yujing Zhang}, \bibinfo{person}{Zhe Wang}, \bibinfo{person}{Lejian Ren}, \bibinfo{person}{Ying Fan}, \bibinfo{person}{Xiaoqiang Zhu}, {and} \bibinfo{person}{Kun Gai}.} \bibinfo{year}{2020}\natexlab{}.
\newblock \showarticletitle{Search-based user interest modeling with lifelong sequential behavior data for click-through rate prediction}. In \bibinfo{booktitle}{\emph{Proceedings of the 29th ACM International Conference on Information \& Knowledge Management}}. \bibinfo{pages}{2685--2692}.
\newblock


\bibitem[Schnabel et~al\mbox{.}(2016)]%
        {ipw}
\bibfield{author}{\bibinfo{person}{Tobias Schnabel}, \bibinfo{person}{Adith Swaminathan}, \bibinfo{person}{Ashudeep Singh}, \bibinfo{person}{Navin Chandak}, {and} \bibinfo{person}{Thorsten Joachims}.} \bibinfo{year}{2016}\natexlab{}.
\newblock \showarticletitle{Recommendations as treatments: Debiasing learning and evaluation}. In \bibinfo{booktitle}{\emph{international conference on machine learning}}. PMLR, \bibinfo{pages}{1670--1679}.
\newblock


\bibitem[Su et~al\mbox{.}(2024)]%
        {ddpo}
\bibfield{author}{\bibinfo{person}{Hongzu Su}, \bibinfo{person}{Lichao Meng}, \bibinfo{person}{Lei Zhu}, \bibinfo{person}{Ke Lu}, {and} \bibinfo{person}{Jingjing Li}.} \bibinfo{year}{2024}\natexlab{}.
\newblock \showarticletitle{DDPO: Direct Dual Propensity Optimization for Post-Click Conversion Rate Estimation}. In \bibinfo{booktitle}{\emph{Proceedings of the 47th International ACM SIGIR Conference on Research and Development in Information Retrieval}}. \bibinfo{pages}{1179--1188}.
\newblock


\bibitem[Wang et~al\mbox{.}(2022)]%
        {escm2}
\bibfield{author}{\bibinfo{person}{Hao Wang}, \bibinfo{person}{Tai-Wei Chang}, \bibinfo{person}{Tianqiao Liu}, \bibinfo{person}{Jianmin Huang}, \bibinfo{person}{Zhichao Chen}, \bibinfo{person}{Chao Yu}, \bibinfo{person}{Ruopeng Li}, {and} \bibinfo{person}{Wei Chu}.} \bibinfo{year}{2022}\natexlab{}.
\newblock \showarticletitle{ESCM2: Entire Space Counterfactual Multi-Task Model for Post-Click Conversion Rate Estimation}. In \bibinfo{booktitle}{\emph{Proceedings of the 45th International ACM SIGIR Conference on Research and Development in Information Retrieval}}. \bibinfo{publisher}{ACM}, \bibinfo{pages}{363–372}.
\newblock
\urldef\tempurl%
\url{https://doi.org/10.1145/3477495.3531972}
\showDOI{\tempurl}


\bibitem[Xu et~al\mbox{.}(2022)]%
        {ukd}
\bibfield{author}{\bibinfo{person}{Zixuan Xu}, \bibinfo{person}{Penghui Wei}, \bibinfo{person}{Weimin Zhang}, \bibinfo{person}{Shaoguo Liu}, \bibinfo{person}{Liang Wang}, {and} \bibinfo{person}{Bo Zheng}.} \bibinfo{year}{2022}\natexlab{}.
\newblock \showarticletitle{Ukd: Debiasing conversion rate estimation via uncertainty-regularized knowledge distillation}. In \bibinfo{booktitle}{\emph{Proceedings of the ACM Web Conference 2022}}. \bibinfo{pages}{2078--2087}.
\newblock


\bibitem[Zhang et~al\mbox{.}(2020)]%
        {multiipw}
\bibfield{author}{\bibinfo{person}{Wenhao Zhang}, \bibinfo{person}{Wentian Bao}, \bibinfo{person}{Xiao-Yang Liu}, \bibinfo{person}{Keping Yang}, \bibinfo{person}{Quan Lin}, \bibinfo{person}{Hong Wen}, {and} \bibinfo{person}{Ramin Ramezani}.} \bibinfo{year}{2020}\natexlab{}.
\newblock \showarticletitle{Large-scale causal approaches to debiasing post-click conversion rate estimation with multi-task learning}. In \bibinfo{booktitle}{\emph{Proceedings of The Web Conference 2020}}. \bibinfo{pages}{2775--2781}.
\newblock


\bibitem[Zhou et~al\mbox{.}(2018)]%
        {din}
\bibfield{author}{\bibinfo{person}{Guorui Zhou}, \bibinfo{person}{Xiaoqiang Zhu}, \bibinfo{person}{Chenru Song}, \bibinfo{person}{Ying Fan}, \bibinfo{person}{Han Zhu}, \bibinfo{person}{Xiao Ma}, \bibinfo{person}{Yanghui Yan}, \bibinfo{person}{Junqi Jin}, \bibinfo{person}{Han Li}, {and} \bibinfo{person}{Kun Gai}.} \bibinfo{year}{2018}\natexlab{}.
\newblock \showarticletitle{Deep Interest Network for Click-Through Rate Prediction}. In \bibinfo{booktitle}{\emph{ACM SIGKDD Conference on Knowledge Discovery and Data Mining (KDD)}}.
\newblock


\bibitem[Zhu et~al\mbox{.}(2023)]%
        {dcmt}
\bibfield{author}{\bibinfo{person}{Feng Zhu}, \bibinfo{person}{Mingjie Zhong}, \bibinfo{person}{Xinxing Yang}, \bibinfo{person}{Longfei Li}, \bibinfo{person}{Lu Yu}, \bibinfo{person}{Tiehua Zhang}, \bibinfo{person}{Jun Zhou}, \bibinfo{person}{Chaochao Chen}, \bibinfo{person}{Fei Wu}, \bibinfo{person}{Guanfeng Liu}, {et~al\mbox{.}}} \bibinfo{year}{2023}\natexlab{}.
\newblock \showarticletitle{DCMT: A Direct Entire-Space Causal Multi-Task Framework for Post-Click Conversion Estimation}. In \bibinfo{booktitle}{\emph{2023 IEEE 39th International Conference on Data Engineering (ICDE)}}. IEEE, \bibinfo{pages}{3113--3125}.
\newblock


\end{thebibliography}
\end{document}